# Low-dimensional controllability of brain networks

Remy Ben Messaoud[1], Vincent Le Du[2], Brigitte Charlotte Kaufmann[3], Baptiste Couvy-Duchesne[1,4], Lara Migliaccio[2], Paolo Bartolomeo[3], Mario Chavez[1], Fabrizio De Vico Fallani[1,*]

[1]Sorbonne Université, Paris Brain Institute-ICM, CNRS, Inria, Inserm, AP-HP, Hôpital de la Pitié-Salpêtrière, F-75013, Paris, France.

[2]FrontLab, INSERM U1127, Paris Brain Institute-ICM, AP-HP, Hôpital de la Pitié-Salpêtrière, F-75013, Paris, France.

[3]PicnicLab, INSERM U1127, Paris Brain Institute-ICM, AP-HP, Hôpital de la Pitié-Salpêtrière, F-75013, Paris, France.

[4]Institute for Molecular Bioscience, the University of Queensland, St Lucia, QLD, Australia.

* Corresponding author

**Email:** fabrizio.de-vico-fallani@inria.fr

**Abstract**
Network controllability is a powerful tool to study the causal relationships in complex systems and identify the driver nodes for steering the network dynamics into desired states.
However, due to ill-posed conditions, results become unreliable when the number of drivers becomes too small compared to the network size. This is a very common situation, particularly in real-world applications, where the possibility to access multiple nodes at the same time is limited by technological constraints, such as in the human brain.
Although targeting small network parts might improve accuracy in general, challenges may still remain for extremely unbalanced situations, when for example there is one single driver. To address this problem, we developed a mathematical framework that combines concepts from spectral graph theory and output controllability. Instead of controlling the original network dynamics, we aimed to control its low-dimensional embedding into the topological space derived from the Laplacian network structure.
By performing extensive simulations on synthetic networks, we showed that a relatively low number of projected components is enough to improve the overall control accuracy, notably when dealing with very few drivers. Based on these findings, we introduced alternative low-dimensional controllability metrics and used them to identify the main driver areas of the human connectome obtained from N=6134 healthy individuals in the UK-biobank cohort. Results revealed previously unappreciated influential regions compared to standard controllability approaches, enabled to draw control maps between distinct specialized large-scale brain systems, and yielded an anatomically-based understanding of cerebral specialization.
Taken together, our results offered a theoretically-grounded solution to deal with network controllability in real-life applications and provided insights into the causal interactions of the human brain.

## Introduction

Network controllability refers to the possibility of identifying the driver nodes in an interconnected system and opportunely modifying their dynamics to steer the system's state towards desired configurations[1]. Network controllability stems from other established disciplines, such as control and graph theory, and has recently witnessed significant theoretic advances. Analytical approaches allowed to establish the impact of network topology on controllability by unveiling the link between the minimum number of drivers necessary to control a network and its node degree distribution[2,3] as well as its eigenvalues' geometric multiplicity[4]. More recently, the use of controllability metrics derived from Gramian-based energy minimization provided analytical tools to measure the potential of candidate drivers to physically control networks[5–7].

The number of applications studying the controllability of real systems has flourished in the last years. By modeling how information is stipulated to pass along the underlying network structure, controllability informs on the dynamical properties of the nodes that cannot be merely obtained by looking at their connectivity alone. Structural controllability can quantify the ability to manipulate and regulate the flow in power grids[8] or transportation networks[9], influence information and behavior in social networks[10], and gain insights into complex biological processes, such as gene expression regulation[11] and brain functions[12–14].

Despite its great potential, network controllability's concrete impact remains quite limited due to the difficulty of identifying in a reliably not only the number of drivers but more importantly where they are located and what is the control signal that is to be injected to push the system into desired states. This is a typical computationally expensive problem, further aggravated by the extremely large condition numbers that arise when the driver nodes are less than one-fifth of the entire network[15]. In those very common situations, round-off numerical errors arise and the network cannot be controlled in practice despite placing the drivers strategically to ensure theoretical controllability[6,15,16]. Notably in neuroscience, studying the control influences between differently specialized brain systems is not only important from a fundamental perspective (endogenous control)[17] but often a necessity imposed by the existing technology that allows to stimulate only one or few regions at a time (exogenous control)[18,19]. However, because of the aforementioned problems, the typical controllability metrics are often close to zero or negative, making it difficult to predict the actual impact on brain functioning and resulting behavior[20–22].

To address these limitations disparate approaches have been introduced. From an algorithmic perspective, using a stepwise exploration to measure the extent of the network that can be controlled by a driver, provides some benefit on the overall accuracy[23]. Targeting specific parts of the system is perhaps the most intuitive way to lower the imbalance between the number of drivers and the size of the network to control[24,25]. In general, output controllability can be adopted to reduce the problem dimensionality by controlling condensed components of the original network state. For example, controlling the averaged state of the nodes within separate groups, or clusters, has been demonstrated to be an effective strategy to improve precision[26].

However, average-state controllability is an over-approximation of the real dynamics as nodes in the same cluster might have very different functional states. Because network dynamics are constrained by the underlying structure[27,28], one possibility would be instead to exploit the existing connectivity to have a more representative aggregation. By leveraging spectral graph theory, we introduced a framework that embeds the original network state into few representative components characterizing the structure of the system at different organizational scales[29].

We formally introduced our mathematical approach for linear-time invariant systems and evaluated via extensive simulations on synthetic networks the control precision and representativeness in contrast to the standard approaches. Eventually, by projecting the Gramian into a low-dimensional space we introduced a parsimonious controllability metric to identify the driver nodes of a network. We used such low-dimensional controllability metric to identify the most influential areas of the human connectome obtained by gathering more than 6000 samples from the UK-biobank cohort[30] and provide a map of the causal interactions between functional brain systems and characterize cerebral specialization.

# Results

Without loss of generality, we considered unweighted and undirected networks whose dynamics are governed by the following linear time-invariant equation

$$\dot{x}(t) = Ax(t) + Bu(t)$$
$$y(t) = Cx(t) \quad (1)$$

Where $x \in \mathbb{R}^n$ contains the states of the $n$ nodes, $A \in \mathbb{R}^{n \times n}$ is a stabilized version of the network adjacency matrix $G$ and $B \in \mathbb{R}^{n \times n_d}$ directs the external input $u \in \mathbb{R}^{n_d}$ into $n_d \leq n$ distinct driver nodes. Here, $y \in \mathbb{R}^r$ with $r < n$ represents the low-dimensional output state obtained by projecting the entire state vector $x$ into fewer components via the so-called output matrix $C \in \mathbb{R}^{r \times n}$ [31,32].

In general, different configurations for the output matrix can be chosen to aggregate the original network state into fewer components. Instead of selecting $C$ in an arbitrary or predetermined fashion, we derived it directly from the network structure by considering the Laplacian linear operator $L = D - A$, where $D$ is a diagonal matrix containing the node degree sequence[29]. By construction, $L$ can be decomposed into several eigenvectors $V = [V_1, V_2, ..., V_n]$ or eigenmaps, associated with incrementing eigenvalues that inform on the network structure load from coarser to finer-grained topological scales. Eigenmaps serve as a basis to derive a spectral representation of the network state $\tilde{x} = V^T x$, here referred to as *eigenstate* (**Fig 1a**). By selecting a smaller number of eigenmaps, we aimed to control the low-dimensional eigenstate

$$y(t) = H_r \tilde{x}(t) = C^{EIG} x(t) \quad (2)$$

where the output matrix $C^{EIG} = H_r V^T$ is obtained by selecting and reordering a number $r < n$ of eigenmaps via a filtering matrix $H_r$ (**Fig 1b**).

## Tradeoff between control precision and representativeness

We considered hierarchical modular small-world (HMSW) networks whose Laplacian eigenmaps informed on the network partition across multiple scales[33] (**Fig 2a**, **Materials and Methods**). Without loss of generality[34], we set the initial network state in the origin $x_0 = \vec{0} \in \mathbb{R}^n$ and the final state as a random departure from it $x_f \sim \mathcal{N}(\mu_f = 1, \sigma_f^2 = 100) \in \mathbb{R}^n$, which corresponded to steering the related eigenstate from $y_0 = \vec{0}$ to $y_f = C^{EIG} x_f$. Candidate drivers were chosen by ranking the nodes according to their betweenness centrality so as to ensure homogeneous distances from the rest of the network[35]. We then computed the associated input signals $u(t)$ by minimizing the output cost function $J_\rho(u, t_f) = \| y_f - y(t_f) \|^2 + \rho \int_0^{t_f} \| u(\tau) \|^2 d\tau$, where $\rho$ is a regularization parameter weighting the accuracy of the solution with respect to the signal energy[36] (**Text S1**). By injecting the optimal inputs back into the model dynamics, we considered the cosine similarity[37] to measure the control accuracy in terms of *precision* $\delta = \frac{y(t_f) \cdot y_f}{\|y(t_f)\| \|y_f\|}$. Here, $\rho = 10^{-4}$ and $t_f = 1$ ensured sufficiently accurate solutions (**Fig S1a**). To evaluate the impact of the low-dimensional control on the original network state, we also measured control accuracy in terms of *representativeness* $\eta = \frac{x(t_f) \cdot x_f}{\|x(t_f)\| \|x_f\|}$.

Results showed that the number of eigenmaps significantly affects the control precision. The lower was $r$, the higher was $\delta$ regardless of the number of drivers $n_d$. However, having more drivers led to higher precision and faster transitions towards the largest values (**Fig 2b**). To simulate more realistic situations, we considered the case of single-driver control. On average, reducing the number of eigenmaps improved the control precision but worsened representativeness. Optimal tradeoffs could be then explored considering unbiased indicators such as the total accuracy $\delta + \eta$ (**Fig 2c**). We next targeted the control of a subnetwork $S$ with $m \leq n$ nodes and computed the Laplacian taking into account its internal connectivity (**Fig 2d, Materials and Methods**). As in full network control, we computed the related low-dimensional eigenstates and measured the total accuracy as a function of the target size. Results showed an overall benefit in controlling smaller parts of the network, especially when considering internal drivers and using a relatively low number of eigenmaps (**Fig 2e**).

Note that these findings were obtained by selectively including eigenmaps from the most to the least informative in terms of $y$ magnitude. Changing the inclusion criterion, such as ranking them according to the eigenvalues, led to similar tendencies albeit with less representativeness (**Fig S1b**). Notably, results stayed comparable even when considering networks with different topologies, i.e., random and scale-free (**Fig S1c**).

**Low-dimensional controllability of brain systems**
To quantify the ability of individual brain regions to influence the activity of other areas, we introduced a control metric based on a low-dimensional projection of the Gramian matrix [5]

$$\mathcal{W} = \int_0^\infty e^{A\tau} BB^T e^{A^T\tau}\, d\tau \qquad (3)$$

where $B$ selects the i[th] candidate driver. Similar to **Eq. 2**, we operated such projection via the $C^{EIG}$ matrix, i.e., $\mathcal{W}^{EIG} = C^{EIG} \mathcal{W} C^{EIG^T}$. Its smallest eigenvalue $\lambda_{min}^{EIG}$ gave a low-dimensional version of the worst-case control centrality $\lambda_{min}$, measuring the amount of energy needed for node $i$ to steer the network into the most difficult-to-reach state[38].

We considered structurally weighted brain networks obtained from healthy human subjects in the UK-biobank cohort. Brain networks consisted of $n$=214 nodes including cortical and subcortical regions of the Schaefer atlas[39] (**File S1**). They were grouped into nine differently specialized functional systems[40] (**Fig 3a**, **Materials and Methods).** Our first main result showed that reducing the number of eigenmaps allowed the entire brain to be controllable from every single node (i.e., $\lambda_{min}^{EIG} > 0$) whereas this failed when using standard $\lambda_{min}$. The transition to full controllability occurred at $r^* \leq 5$ in line with previous evidence[15] and used in all our subsequent analyses (**Fig 3b**). Compared to standard worst-case controllability some regions exhibited similar relative spatial contributions (e.g. SUB), while others faded out or gained importance notably in the visual-temporal part of the right hemisphere (e.g. VIS, DAN) (**Fig 3c**).

Using the target controllability approach described in the previous section, we next considered the Laplacian of the subnetworks corresponding to the brain systems and evaluated their controllability (**Fig S2**). Results pooled from all subjects showed that systems could be controlled from both internal and external nodes enabling to derive optimal driver-target configurations (**Fig 4a**, **Tab 1**). More in general, we reported a system controllability $\langle \lambda_{min}^{EIG} \rangle$ hierarchy with subcortical (SUB) and visual (VIS) ones among the easiest to control (**Fig 4b**). Internal drivers exhibited higher control centrality compared to external ones, indicating a preferential system self-regulation $\langle \lambda_{min}^{EIG} \rangle_{in}$ with respect to external regulation $\langle \lambda_{min}^{EIG} \rangle_{out}$ (**Materials and methods**). That was particularly evident for SUB, but also for the primary systems VIS and somatomotor (SMN), the latter to a lesser extent (**Fig 4c**). This result was not necessarily due to the spatial proximity between the driver and the target, but rather to the topological distance in terms of shortest paths (**Fig S2**).

We then investigated the system lateralization in terms of self-regulation to accumulate in one or the other hemisphere. For each targeted system, we computed a lateralization index $\varphi = \frac{\zeta^R - \zeta^L}{\zeta^R + \zeta^L}$, where $\zeta = \langle \lambda_{min}^{EIG} \rangle_{in}$ for the nodes in the right (R) or left (L) hemisphere. Results showed significant system-dependent lateralization, which was not merely due to the possible differences in terms of system size between hemispheres (**Fig 4d**). Such lateralization was perfectly in line with the known functional experimental evidence, such as the right-dominated limbic and dorsal-attention network, and could not be retrieved when using the standard worst-case control metric $\lambda_{min}$.

**Mapping inter-system controllability in the brain**
To better understand how different systems were influencing each other we next studied their reciprocal role as drivers and targets. For each pair of systems *i* and *j* we established a directed weighted link by calculating the geometric mean control centrality of the nodes in *i* when targeting *j*, i.e., $\langle \lambda_{min}^{EIG} \rangle_{i \to j}$. By inspecting the resulting meta-graph, we found a heterogeneous strength and distribution of causal influences that enabled to identify the preferential targets and drivers for each system (**Fig 5a**). Notably, primary systems exerted a strong controlling influence over the dorsal-attentional network (VIS→DAN, SMN→DAN), while more balanced interactions were found between associative systems (e.g., FPCN↔DMN, DMN↔LIM).

To quantify the control unbalance, we computed for each system the difference between the outgoing and incoming sum of the weighted links of the meta-graph. Results confirmed that both VIS and SMN exhibited a significantly higher tendency to control as compared to SUB and DAN which instead appeared much easier to be controlled (**Fig. 4b**). A more balanced profile was instead observed for the other systems, such as the limbic system (LIM). Similar results were also obtained when using a finer-grained version of the Yeo2011 parcellation (**Fig. S4**). We finally investigated hemispheric dominance in terms of control capacity. To do so, we computed the $\lambda_{min}^{EIG}$ value of each node when targeting the two hemispheres, separately. We defined *i)* the *ipsilateral* control as the mean $\langle\lambda_{min}^{EIG}\rangle_{i\rightarrow i}$ of the nodes in one hemisphere targeting the same hemisphere, and *ii)* the *contralateral* control as the mean $\langle\lambda_{min}^{EIG}\rangle_{i\rightarrow j}$ of the same nodes targeting the other hemisphere.

Taken separately, each hemisphere exhibited a stronger capacity to control itself in contrast to its ability to influence the other. However, non-trivial results emerged when comparing the two hemispheres. While the ipsilateral control of the left hemisphere (L→L) was significantly higher than the right hemisphere (R→R), the contralateral control of the right hemisphere (R→L) was significantly higher than the left hemisphere (L→R) (**Fig 5c**). This global behavior could be explained by the same tendency across all the local brain systems, but the temporoparietal junction (TPJ) whose contralateral control exhibited an opposite trend (**Fig 5d**). Altogether, these findings indicated that the left hemisphere exhibits better self-regulation, while the right hemisphere has a stronger ability to influence its counterpart.

## Discussion
### Spectral graph theory and network controllability
Spectral graph theory (SGT) is a powerful and versatile framework for understanding and analyzing complex interconnected systems. SGT is particularly concerned with the eigenvalues and eigenvectors of the Laplacian matrix associated with a graph and informs on the global and finer-grained connectivity structures in the network[41,42]. Laplacian eigenvectors, also known as eigenmaps, find various applications in network science, such as data clustering and dimensionality reduction via spectral embedding[43]. From a broader graph signal processing angle, the Laplacian eigenmaps technique informs on dynamical processes like synchronization or diffusion and is a suitable framework to analyze how external perturbations propagate in a network[27]. More recently, SGT has been developed to identify connectivity gradients in the brain[44], disentangle brain structure and function[45], improve machine learning for systems neuroscience[46], and decompose brain dynamics into graph Fourier modes[47,48].

Despite its potential to reduce the complexity of signals on graphs, SGT has been poorly explored in the context of network controllability. Network controllability provides a mathematically grounded framework to identify drivers, influence network dynamics and understand causal relationships, but it still faces critical challenges that hamper its usability in many practical situations[5,15,16]. This is mainly due to the presence of ill-posed conditions that make the problem numerically unstable with important consequences on the solutions, such as an overestimation of the drivers or completely unreliable input-controlling signals[15]. Hence, the development of methods to reduce the problem complexity is paramount to enhance the usability and interpretation of network controllability in real-world applications. Targeting specific subnetworks[23–25] or reducing the network to fewer averaged nodes[26,49] have shown substantial benefits, yet they represent approximations that oversimplify the original network structure.

SGT offers a more flexible solution to reduce problem complexity enabling to write the network states as a linear combination of its actual connectivity structures (the eigenmaps) across different topological scales. By establishing a link between SGT and output controllability, we showed that controlling the most representative spectral components from the graph Fourier transform, instead of the original network state, leads to significant improvements even in the presence of unbalanced $(^{n_d}/_m \ll 1)$ and noisy configurations. The price to pay for such a control precision consisted of an almost equivalent loss in the representativeness of the solution in terms of capacity to reproduce the original network state. Optimal trade-offs can be further investigated in future research to better capture the interplay between network topology, the number of output components, and the smoothness of the state solution.

**Theoretical and effective controllability of brain networks**
Network controllability can provide insights into how different regions of the brain interact to perform various cognitive and motor functions by passing electrical activity through white matter connections[14,50]. Its theoretical framework can be used to experimentally test basic questions in network neuroscience and inform noninvasive brain stimulation or neurofeedback approaches to enhance cognitive functions, such as memory and learning[51]. Furthermore, investigating the controllability of brain networks can shed light on the causal mechanisms underlying neurological[52] and psychiatric disorders[53] as well as help identify potential entry points for therapeutic interventions, such as brain stimulation techniques for conditions like epilepsy, Parkinson's disease, stroke and chronic pain[1].

Despite its potential, brain network controllability faces a number of methodological challenges that limit its usability. This is mainly due to the unreliability of control centrality metrics, such as the worst-case control $\lambda_{min}$, which are mathematically correct but have little translational value when the number of drivers is much lower than the nodes to control. The main consequence is that is not always possible to infer the control power of single brain areas because it cannot be assessed numerically and remains difficult to interpret[20–22]. Here, we showed that low-dimensional controllability allows overcoming this limitation and reveals previously unappreciated brain drivers in the VIS and DAN that could not be identified by merely looking at their connectivity strength (**Fig 3, Fig S5**). SUB was the most controllable system possibly due to its rich-club organization integrating information from remote parts of the brain[56], and together with VIS and SMN exhibited a high tendency to self-regulate as compared to other associative systems[57].

The analysis of the self-regulation's lateralization allowed to specify the known functional[58] and anatomical[59] hemispheric asymmetries such as the right preference for TPJ and SVAN partially overlapping with the ventral attention network, and for LIM in line with the control of emotional processing[60]. In addition, the found left-hemisphere advantage of SMN and DMN aligned with the suggested preferential role in action selection[61] and endogenously generated cognition[62]. More importantly, we provided new insights for those systems whose asymmetry is more controversial such as VIS[63] and DAN[64,65]. From a control-theoretic perspective, these systems exhibited a strong tendency to self-regulate via the right hemisphere and provided a novel anatomical basis for the common clinical observation that spatial neglect is more frequent, persistent, and severe after damage to the right hemisphere as compared to lesions to the left homotopic areas[66].

**Causal relationships between brain systems**
The presence of several systems, often called "networks", coexisting in the brain is an emerging feature of its intrinsic functioning and crucial to understanding human behavior in both healthy and pathological conditions[58]. By leveraging connectomics to trace functional connectivity, several large-scale brain systems have been identified that are important, among others, for decision making (FPCN), emotion, motivation and memory (LIM), attention for externally-directed tasks (DAN), introspective thoughts (DMN), visual and somatosensory processing (VIS, SMN), and internal/external thinking (SVAN)[67]. Notably, these systems exhibit a hierarchy to their operation, but they do not work in isolation. Instead, they are known to integrate and synchronize to carry out complex functions[17,40].

At a larger scale, the brain hemispheres can also be considered as anatomically separated systems that functionally interact to accomplish complex tasks[68–70]. Cerebral specialization is not only important to underpin the neural basis of language, attention, and motor control, but it is also critical to assess and treat related deficits due to mental or neurological disorders[68,71]. Yet, determining how large-scale systems simultaneously influence each other in a causal manner is still poorly understood. Here, we showed that drivers in the primary systems (VIS, SMN) were particularly apt to influence distributed activity in associative and attentional systems (e.g., DAN, SVAN), while drivers in the prefrontal cortex exhibited facilitated control of SUB,

SMN, and DMN (**Fig 4**). This result was further confirmed by the meta-wiring diagram showing how brain systems simultaneously interact from a control theoretic perspective.

The role of primary systems as controlling drivers and that of associative systems as controlled targets, match previous results obtained with network communication models of brain cognition[72] and align with theories postulating that the brain topography is organized along spatial and functional gradients[73,74]. These gradients span from peripheral regions that handle perception and action to core areas dealing with more abstract cognitive functions. Our findings revealed that core attentional networks are in particular controlled by primary peripheral systems, which are instead more influenced by external events. In terms of hemispheric asymmetries, our results extend recent evidence using fMRI functional connectivity suggesting a preference for the left hemisphere to interact with itself and for the right hemisphere to integrate information both locally and from the other hemisphere[75]. Our findings provide an anatomical basis for this fundamental result by unveiling unbalanced directed influences within and between hemispheres.

**Conclusion**
Network controllability is not only crucial to identify driver nodes, but in general to explore causal relationships in complex interconnected systems such as the human brain. By leveraging principles from spectral graph theory, the introduced low-dimensional controllability framework offers a reliable tool to interrogate, influence and ultimately understand the behavior of complex systems.

## Materials & Methods

**Laplacian of stabilized networks.** The graph Laplacian $L = D - A$ is a symmetric matrix that can be rewritten via singular value decomposition as $L = V^T \Lambda V$, where $\Lambda$ contains its increasing eigenvalues ($\lambda_1 \leq \lambda_2 \ldots \leq \lambda_n$) and $V = [V_1, V_2, \ldots, V_n]$ are the associated eigenvectors[8,29]. By construction, eigenvectors are orthonormal (i.e., $V^T V = I$) and constitute a projection basis to obtain the spectral representation of any signal $x$ on a graph via the so-called Graph Fourier Transform (GFT)[29], i.e., $\tilde{x} = V^T x$. Here, networks corresponded to graphs whose adjacency matrices $G$ coded for the presence/weight of links between the nodes. To ensure the stability of the system and the existence of the Gramian we next added negative self-loops to the network via a linear transformation $A = G - cI$, with the constant $c$ chosen here as the maximum eigenvalue of $G$[31,76]. It can be easily demonstrated that such transformation does not alter the network Laplacian which is a linear operator (**Text S1**).

**Target network control.** Let $S = [s_1, s_2, \ldots, s_m]$ be a subset of $m \leq n$ target nodes indexed by $s_i \in [1, n]$ and $G(S, S)$ the corresponding adjacency matrix. We formulated the related target control problem by considering the Laplacian of the subnetwork $L_S = D_S - G(S, S)$, where $D_S$ is a diagonal matrix that contains the degree sequence of the target nodes excluding the links to the rest of the network. Note that a precise relation holds with the submatrix of the original Laplacian $L(S, S) = L_S + Z_S$, where $Z_S$ is a diagonal matrix containing the sum of the links connecting the target nodes to the rest of the network[77]. Hence, $L_S$ exactly matches the actual spatial topological scales of the subnetwork and it is not biased by external influences. By considering the state of the subnetwork $x_S(t)$ and the related eigenmaps $V_S$ we obtained the target *eigenstate* $\tilde{x}_S(t) = V_S^T x_S(t)$ and derived its low-dimensional output

$$y_S^{EIG}(t) = H_r \tilde{x}_S(t) = C_S^{EIG} x_S(t) \qquad (4)$$

where the output matrix $C_S^{EIG} = H_r V_S^T$ is obtained by selecting and reordering a number $r < m$ of spectral components via a filtering matrix $H_r$.

**Hierarchical modular small-world network and controllability metrics.** We used the HMSW model[33] to generate synthetic networks with $n = 256$ nodes, $8 = \log_2(256)$ hierarchical levels, and an average network density of 0.035. Other parameters were: initial cluster size=2, and connection density fall-off per level=2.5. The choice of this model and its parameters was particularly relevant for the purposes of this study and has been widely adopted in topological analysis of brain networks[33] which exhibit hierarchical modularity[78].

Starting from the worst-case low-dimensional control centrality $\lambda_{min}^{EIG}$ of each node, we defined aggregated metrics to measure larger topological effects, i.e., i) *system controllability* $\langle\lambda_{min}^{EIG}\rangle$ as the mean of all the nodes targeting a specific subnetwork, ii) *self-regulation* $\langle\lambda_{min}^{EIG}\rangle_{in}$ as the mean of all the nodes inside a target subnetwork, and iii) *external regulation* $\langle\lambda_{min}^{EIG}\rangle_{out}$ as the mean of all the nodes outside a target subnetwork.

**Neuroimaging data and tractography.** The UK-biobank dataset is a large-scale cohort containing clinical, genetic, and imaging data[30]. We selected N=6134 human subjects who had both T1 weighted and diffusion MRI and no known disease history, i.e., international classification of diseases ICD-10='none'. Population statistics: 50.73% women, 88.72% right-handed, and age 62.41±7.25 at the time of the MRI scans. Informed consent was obtained from all UK Biobank participants. We rejected those who requested the withdrawal of their data. Procedures are controlled by a dedicated Ethics and Guidance Council (http://www.ukbiobank.ac.uk/ethics), with the Ethics and Governance Framework available at https://www.ukbiobank.ac.uk/media/0xsbmfmw/egf.pdf. IRB approval was also obtained from the North West Multi-centre Research Ethics Committee. This research has been conducted using the UK Biobank Resource under Application Number 53185. Downloaded imaging data were corrected and preprocessed with the UK-biobank pipeline[79].

**Tractography and connectome construction**. As for the processing, we computed Tissue response and Fiber Orientation Distribution using multi-tissue and multi-shell algorithms in MRtrix3 [80]. T1 images were aligned to an extracted mean b0 volume via the FLIRT function in FSL[81]. We performed a 5-tissue type segmentation to compute the grey-white matter interface. These were then used in MRtrix3 to compute an anatomically constrained tractography with a cut-off of 0.1 and a density of 1M streamline that was shown to be sufficient for reproducibility[82]. As for the parcellation, we used the Schaefer atlas of 200 cortical regions (i.e., nodes) exhibiting high structural and functional representativeness[83]. Its regions were grouped into eight cortical systems according to the Yeo2011 mapping[40]: VIS, SMN, DAN, SVAN, LIM, FPCN, DMN, TPJ. The cortical atlas was complemented by 14 subcortical regions from the FreeSurfer segmentation that form the network SUB[57]. The parcellation was transferred to the subject space using the T1 linear co-registration and the UK-biobank warp field. It was finally dilated and masked to be used in MRtrix3 along the SIFT[84] for the connectome extraction. Fiber assignment was done with a radial search of 3mm and the resulting connectomes were symmetric with a zero diagonal. Link weights correspond to the number of fibers between two nodes.

**Reproducibility**

All the simulations for both synthetic and brain networks were done in MATLAB and the code is made available at github.com/Inria-NERV/Network-Control. The HMSW model and shortest paths were computed with the BCT toolbox (bctnet.com)[85]. All neuroimaging data are available upon request from the UK-biobank (ukbiobank.ac.uk) and the code of the processing pipeline is available at github.com/UKB-dwi-2-connectome.

**Acknowledgments**
This work was supported by the European Research Council (ERC) under the European Union's Horizon 2020 research and innovation program (Grant Agreement N° 864729). BCD is supported by a CJ Martin Fellowship (NHMRC app 1161356). The authors declare no competing interests.

**References**

1. Liu, Y.-Y. & Barabási, A.-L. Control principles of complex systems. *Rev. Mod. Phys.* **88**, 035006 (2016).
2. Liu, Y.-Y., Slotine, J.-J. & Barabási, A.-L. Controllability of complex networks. *Nature* **473**, 167–173 (2011).
3. Nie, S., Wang, X.-W., Wang, B.-H. & Jiang, L.-L. Effect of correlations on controllability transition in network control. *Sci. Rep.* **6**, 23952 (2016).
4. Yuan, Z., Zhao, C., Di, Z., Wang, W.-X. & Lai, Y.-C. Exact controllability of complex networks. *Nat. Commun.* **4**, 2447 (2013).
5. Pasqualetti, F., Zampieri, S. & Bullo, F. Controllability Metrics, Limitations and Algorithms for Complex Networks. *IEEE Trans. Control Netw. Syst.* **1**, 40–52 (2014).
6. Wang, L.-Z., Chen, Y.-Z., Wang, W.-X. & Lai, Y.-C. Physical controllability of complex networks. *Sci. Rep.* **7**, 40198 (2017).
7. Lindmark, G. Minimum energy control for complex networks. *Sci. Rep.* 14 (2018).
8. Yang, D.-S., Sun, Y.-H., Zhou, B.-W., Gao, X.-T. & Zhang, H.-G. Critical Nodes Identification of Complex Power Systems Based on Electric Cactus Structure. *IEEE Syst. J.* **14**, 4477–4488 (2020).
9. Rinaldi, M. Controllability of transportation networks. *Transp. Res. Part B Methodol.* **118**, 381–406 (2018).
10. Cremonini, M. & Casamassima, F. Controllability of social networks and the strategic use of random information. *Comput. Soc. Netw.* **4**, 10 (2017).
11. Rajapakse, I., Groudine, M. & Mesbahi, M. Dynamics and control of state-dependent networks for probing genomic organization. *Proc. Natl. Acad. Sci.* **108**, 17257–17262 (2011).
12. Yan, G. *et al.* Network control principles predict neuron function in the Caenorhabditis elegans connectome. *Nature* **550**, 519–523 (2017).
13. Gu, S. *et al.* Controllability of structural brain networks. *Nat. Commun.* **6**, 8414 (2015).
14. Tang, E. & Bassett, D. S. Colloquium: Control of dynamics in brain networks. *Rev. Mod. Phys.* **90**, 031003 (2018).
15. Sun, J. & Motter, A. E. Controllability Transition and Nonlocality in Network Control. *Phys. Rev. Lett.* **110**, 208701 (2013).
16. Yan, G. *et al.* Spectrum of controlling and observing complex networks. *Nat. Phys.* **11**, 779–786 (2015).
17. Bressler, S. L. & Menon, V. Large-scale brain networks in cognition: emerging methods and principles. *Trends Cogn. Sci.* **14**, 277–290 (2010).
18. Dayan, E., Censor, N., Buch, E. R., Sandrini, M. & Cohen, L. G. Noninvasive brain stimulation: from physiology to network dynamics and back. *Nat. Neurosci.* **16**, 838–844 (2013).
19. Momi, D. *et al.* Network-level macroscale structural connectivity predicts propagation of transcranial magnetic stimulation. *NeuroImage* **229**, 117698 (2021).
20. Tu, C. *et al.* Warnings and Caveats in Brain Controllability. *NeuroImage* **176**, 83–91 (2018).
21. Suweis, S. *et al.* Brain controllability: Not a slam dunk yet. *NeuroImage* **200**, 552–555 (2019).
22. Stocker, J. E., Nozari, E., Vugt, M. van, Jansen, A. & Jamalabadi, H. Network controllability measures of subnetworks: implications for neurosciences. *J. Neural Eng.* **20**, 016044 (2023).
23. Bassignana, G. *et al.* Stepwise target controllability identifies dysregulations of macrophage networks in multiple sclerosis. *Netw. Neurosci.* **5**, 337–357 (2021).
24. Gao, J., Liu, Y.-Y., D'Souza, R. M. & Barabási, A.-L. Target control of complex networks. *Nat. Commun.* **5**, 5415 (2014).
25. Klickstein, I., Shirin, A. & Sorrentino, F. Energy scaling of targeted optimal control of complex networks. *Nat. Commun.* **8**, 15145 (2017).
26. Casadei, G., Canudas-de-Wit, C. & Zampieri, S. Model Reduction Based Approximation of the Output Controllability Gramian in Large-Scale Networks. *IEEE Trans. Control Netw. Syst.* **7**, 1778–1788 (2020).
27. Boccaletti, S., Latora, V., Moreno, Y., Chavez, M. & Hwang, D.-U. Complex networks: Structure and dynamics. *Phys. Rep.* **424**, 175–308 (2006).
28. Porter, M. & Gleeson, J. *Dynamical Systems on Networks: A Tutorial*. vol. 4 (Springer International Publishing, 2016).
29. Chung, F. R. K. *Spectral graph theory*. (Published for the Conference Board of the mathematical sciences by the American Mathematical Society, 1997).
30. Sudlow, C. *et al.* UK Biobank: An Open Access Resource for Identifying the Causes of a Wide Range of Complex Diseases of Middle and Old Age. *PLoS Med.* **12**, e1001779 (2015).

31. Anderson, B. D. O. & Moore, J. B. *Optimal Control: Linear Quadratic Methods*. (Prentice Hall, 1990).
32. Wang, L., Chen, G., Wang, X. & Tang, W. K. S. Controllability of networked MIMO systems. *Automatica* **69**, 405–409 (2016).
33. Sporns, O. Small-world connectivity, motif composition, and complexity of fractal neuronal connections. *Biosystems* **85**, 55–64 (2006).
34. Luenberger, D. G. *Introduction to dynamic systems: theory, models, and applications*. (Wiley, 1979).
35. Duan, C., Nishikawa, T. & Motter, A. E. Prevalence and scalable control of localized networks. *Proc. Natl. Acad. Sci.* **119**, e2122566119 (2022).
36. Shirin, A., Klickstein, I. S. & Sorrentino, F. Optimal control of complex networks: Balancing accuracy and energy of the control action. *Chaos Interdiscip. J. Nonlinear Sci.* **27**, 041103 (2017).
37. Senoussaoui, M., Kenny, P., Stafylakis, T. & Dumouchel, P. A Study of the Cosine Distance-Based Mean Shift for Telephone Speech Diarization. *IEEEACM Trans. Audio Speech Lang. Process.* **22**, 217–227 (2014).
38. Baggio, G., Pasqualetti, F. & Zampieri, S. Energy-Aware Controllability of Complex Networks. *Annu. Rev. Control Robot. Auton. Syst.* **5**, 465–489 (2022).
39. Schaefer, A. *et al.* Local-Global Parcellation of the Human Cerebral Cortex from Intrinsic Functional Connectivity MRI. *Cereb. Cortex N. Y. N 1991* **28**, 3095–3114 (2018).
40. Thomas Yeo, B. T. *et al.* The organization of the human cerebral cortex estimated by intrinsic functional connectivity. *J. Neurophysiol.* **106**, 1125–1165 (2011).
41. Villegas, P., Gili, T., Caldarelli, G. & Gabrielli, A. Laplacian renormalization group for heterogeneous networks. *Nat. Phys.* **19**, 445–450 (2023).
42. Gfeller, D. & De Los Rios, P. Spectral Coarse Graining and Synchronization in Oscillator Networks. *Phys. Rev. Lett.* **100**, 174104 (2008).
43. Shuman, D. I., Narang, S. K., Frossard, P., Ortega, A. & Vandergheynst, P. The emerging field of signal processing on graphs: Extending high-dimensional data analysis to networks and other irregular domains. *IEEE Signal Process. Mag.* **30**, 83–98 (2013).
44. Huntenburg, J. M., Bazin, P.-L. & Margulies, D. S. Large-Scale Gradients in Human Cortical Organization. *Trends Cogn. Sci.* **22**, 21–31 (2018).
45. Preti, M. G. & Van De Ville, D. Decoupling of brain function from structure reveals regional behavioral specialization in humans. *Nat. Commun.* **10**, 4747 (2019).
46. Richiardi, J., Achard, S., Bunke, H. & Van De Ville, D. Machine Learning with Brain Graphs: Predictive Modeling Approaches for Functional Imaging in Systems Neuroscience. *IEEE Signal Process. Mag.* **30**, 58–70 (2013).
47. Lioi, G., Gripon, V., Brahim, A., Rousseau, F. & Farrugia, N. Gradients of connectivity as graph Fourier bases of brain activity. *Netw. Neurosci.* **5**, 322–336 (2021).
48. Huang, W. *et al.* A Graph Signal Processing Perspective on Functional Brain Imaging. *Proc. IEEE* **106**, 868–885 (2018).
49. D'Souza, R. M., di Bernardo, M. & Liu, Y.-Y. Controlling complex networks with complex nodes. *Nat. Rev. Phys.* **5**, 250–262 (2023).
50. Honey, C. J. *et al.* Predicting human resting-state functional connectivity from structural connectivity. *Proc. Natl. Acad. Sci.* **106**, 2035–2040 (2009).
51. Cornblath, E. J. *et al.* Temporal sequences of brain activity at rest are constrained by white matter structure and modulated by cognitive demands. *Commun. Biol.* **3**, 1–12 (2020).
52. Zhou, J., Gennatas, E. D., Kramer, J. H., Miller, B. L. & Seeley, W. W. Predicting regional neurodegeneration from the healthy brain functional connectome. *Neuron* **73**, 1216–1227 (2012).
53. Braun, U. *et al.* Brain network dynamics during working memory are modulated by dopamine and diminished in schizophrenia. *Nat. Commun.* **12**, 3478 (2021).
54. F, F., Y, G., Pe, S., S, S. & Y, Z. Brain controllability distinctiveness between depression and cognitive impairment. *J. Affect. Disord.* **294**, (2021).
55. Scheid, B. H. *et al.* Time-evolving controllability of effective connectivity networks during seizure progression. *Proc. Natl. Acad. Sci. U. S. A.* **118**, undefined-undefined (2021).
56. Bell, P. T. & Shine, J. M. Subcortical contributions to large-scale network communication. *Neurosci. Biobehav. Rev.* **71**, 313–322 (2016).
57. Menardi, A. *et al.* Maximizing brain networks engagement via individualized connectome-wide target search. *Brain Stimul. Basic Transl. Clin. Res. Neuromodulation* **15**, 1418–1431 (2022).

58. Corbetta, M. & Shulman, G. L. Control of goal-directed and stimulus-driven attention in the brain. *Nat. Rev. Neurosci.* **3**, 201–215 (2002).
59. de Schotten, M. T. *et al.* A lateralized brain network for visuospatial attention. *Nat. Neurosci.* **14**, 1245–1246 (2011).
60. Gainotti, G. *Emotions and the Right Side of the Brain*. (Springer International Publishing, 2020).
61. Schluter, N. D., Krams, M., Rushworth, M. F. S. & Passingham, R. E. Cerebral dominance for action in the human brain: the selection of actions. *Neuropsychologia* **39**, 105–113 (2001).
62. Liu, J., Spagna, A. & Bartolomeo, P. Hemispheric asymmetries in visual mental imagery. *Brain Struct. Funct.* **227**, 697–708 (2022).
63. Liu, H., Stufflebeam, S. M., Sepulcre, J., Hedden, T. & Buckner, R. L. Evidence from intrinsic activity that asymmetry of the human brain is controlled by multiple factors. *Proc. Natl. Acad. Sci.* **106**, 20499–20503 (2009).
64. Shulman, G. L. *et al.* Right Hemisphere Dominance during Spatial Selective Attention and Target Detection Occurs Outside the Dorsal Frontoparietal Network. *J. Neurosci.* **30**, 3640–3651 (2010).
65. Nobre, A. C. The attentive homunculus: Now you see it, now you don't. *Neurosci. Biobehav. Rev.* **25**, 477–496 (2001).
66. Lunven, M. & Bartolomeo, P. Attention and spatial cognition: Neural and anatomical substrates of visual neglect. *Ann. Phys. Rehabil. Med.* **60**, 124–129 (2017).
67. Baker, C. M. *et al.* A Connectomic Atlas of the Human Cerebrum-Chapter 1: Introduction, Methods, and Significance. *Oper. Neurosurg. Hagerstown Md* **15**, S1–S9 (2018).
68. Gazzaniga, M. S. Cerebral specialization and interhemispheric communication: Does the corpus callosum enable the human condition? *Brain* **123**, 1293–1326 (2000).
69. Doron, K. W., Bassett, D. S. & Gazzaniga, M. S. Dynamic network structure of interhemispheric coordination. *Proc. Natl. Acad. Sci.* **109**, 18661–18668 (2012).
70. Martínez, J. H., Buldú, J. M., Papo, D., De Vico Fallani, F. & Chavez, M. Role of inter-hemispheric connections in functional brain networks. *Sci. Rep.* **8**, 10246 (2018).
71. Bartolomeo, P. From competition to cooperation: Visual neglect across the hemispheres. *Rev. Neurol. (Paris)* **177**, 1104–1111 (2021).
72. Srivastava, P. *et al.* Models of communication and control for brain networks: distinctions, convergence, and future outlook. *Netw. Neurosci.* **4**, 1122–1159 (2020).
73. Margulies, D. S. *et al.* Situating the default-mode network along a principal gradient of macroscale cortical organization. *Proc. Natl. Acad. Sci. U. S. A.* **113**, 12574–12579 (2016).
74. T. S. Malkinson *et al.* From perception to action: Intracortical recordings reveal cortical gradients of human exogenous attention. *bioRxiv* 2021.01.02.425103 (2022).
75. Gotts, S. J. *et al.* Two distinct forms of functional lateralization in the human brain. *Proc. Natl. Acad. Sci.* **110**, (2013).
76. Karrer, T. M. *et al.* A practical guide to methodological considerations in the controllability of structural brain networks. *J. Neural Eng.* **17**, 026031 (2020).
77. Liu, H., Xu, X., Lu, J.-A., Chen, G. & Zeng, Z. Optimizing Pinning Control of Complex Dynamical Networks Based on Spectral Properties of Grounded Laplacian Matrices. *IEEE Trans. Syst. Man Cybern. Syst.* **51**, 786–796 (2021).
78. Meunier, D., Lambiotte, R. & Bullmore, E. Modular and Hierarchically Modular Organization of Brain Networks. *Front. Neurosci.* **4**, (2010).
79. Alfaro-Almagro, F. *et al.* Image processing and Quality Control for the first 10,000 brain imaging datasets from UK Biobank. *NeuroImage* **166**, 400–424 (2018).
80. Tournier, J.-D. *et al.* MRtrix3: A fast, flexible and open software framework for medical image processing and visualisation. *NeuroImage* **202**, 116137 (2019).
81. Jenkinson, M., Beckmann, C. F., Behrens, T. E. J., Woolrich, M. W. & Smith, S. M. FSL. *NeuroImage* **62**, 782–790 (2012).
82. Roine, T. *et al.* Reproducibility and intercorrelation of graph theoretical measures in structural brain connectivity networks. *Med. Image Anal.* **52**, 56–67 (2019).
83. Luppi, A. I. & Stamatakis, E. A. Combining network topology and information theory to construct representative brain networks. *Netw. Neurosci.* **5**, 96–124 (2021).
84. Smith, R. E., Tournier, J.-D., Calamante, F. & Connelly, A. The effects of SIFT on the reproducibility and biological accuracy of the structural connectome. *NeuroImage* **104**, 253–265 (2015).
85. Rubinov, M. & Sporns, O. Complex network measures of brain connectivity: Uses and interpretations. *NeuroImage* **52**, 1059–1069 (2010).

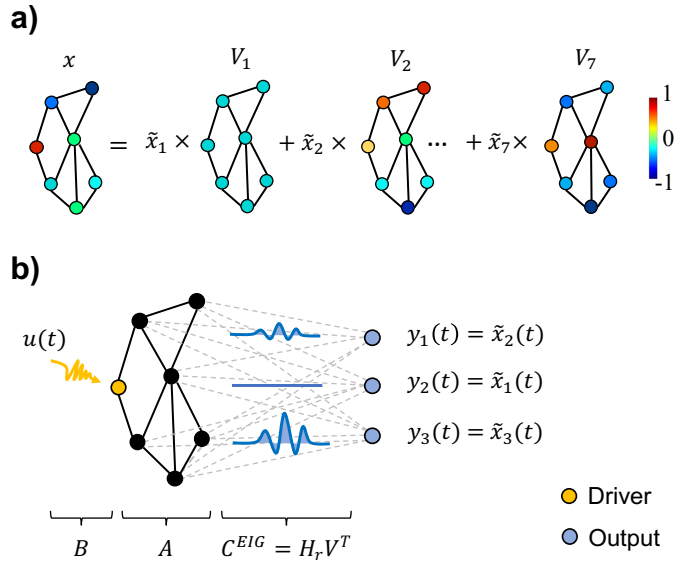

**Figure 1 – Principles of low-dimensional network controllability**

a) Toy network with $n=7$ nodes. Its state $x = [x_1, x_2, \ldots, x_7]$ can be seen as a signal over a graph. By exploiting the Laplacian eigenvectors $V = [V_1, V_2, \ldots, V_7]$ of the network, the signal $x$ can be embedded in a spectral space via the graph Fourier transform (*GFT*) $\tilde{x} = V^T x$. The resulting eigenstate $\tilde{x}$ measures how the signal is spatially distributed across different topological scales, from coarser ($\tilde{x}_1, \tilde{x}_2, \ldots$) to finer-grained ($\ldots, \tilde{x}_6, \tilde{x}_7$) ones. We can also define the inverse operation *iGFT* $x = V\tilde{x}$ that is illustrated in this panel.

b) Low-dimensional controllability in terms of linear-time invariant (LTI) network output control. Instead of focusing on the network state, the goal is to determine the input signal $u(t)$ that steers a low-dimensional output given by the subset of the eigenstate $y(t) = H_r \tilde{x}(t) = C^{EIG} x(t)$, where the output matrix $C^{EIG} = H_r V^T$ is obtained by selecting and reordering a number $r < n$ of spectral components via a filtering matrix $H_r$. In this example, the network has $n=7$ nodes and the first $r = 3$ spectral components are selected and reordered arbitrarily. Here, the filtering matrix $H_r$ is a $3 \times 7$ matrix whose elements are $\{h_{1,2}\} = \{h_{2,1}\} = \{h_{3,3}\} = 1$ and zero elsewhere.

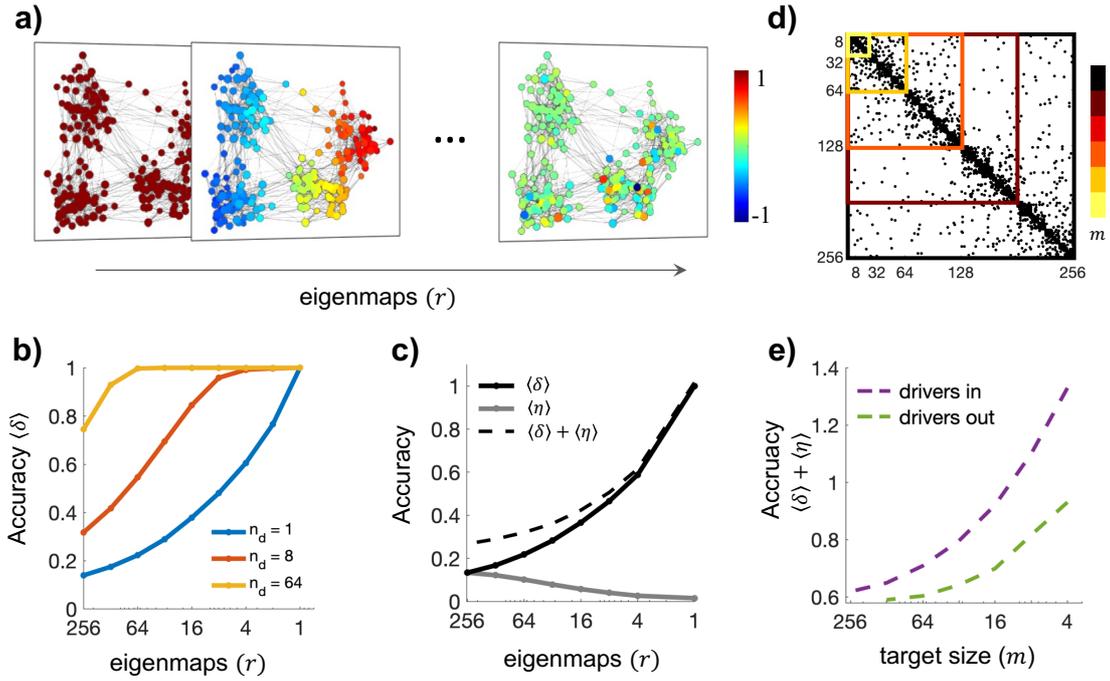

**Figure 2 – Fundamental advantage of low-dimensional controllability**

a) Laplacian eigenvectors (or eigenmaps) for one realization of the hierarchical modular small-world network (HSWM). Colors identify the different spatial contributions of the nodes at increasingly finer topological scales.

b) Control accuracy in terms of precision ($\delta$) as a function of the number of eigenmaps. Different lines correspond to different number of drivers. Candidate drivers were chosen by ranking the nodes according to their betweenness centrality. Results correspond to the mean values obtained from 100 simulated networks.

c) Average node precision $\langle\delta\rangle = \frac{1}{n}\sum_{i=1}^{n}\delta_i$ (black) and representativeness $\langle\eta\rangle = \frac{1}{n}\sum_{i=1}^{n}\eta_i$ (grey) for single-driver control as a function of the number of eigenmaps. Dashed curves correspond to the total accuracy $\langle\delta\rangle + \langle\eta\rangle$. Results correspond to the mean values obtained from 100 simulated networks.

d) Schematic representation of a hierarchical modular small-world network (HSWM) whose target set is progressively expanded by including an increasing number of nodes $m$. The inclusion criterion starts with the nodes in one module and then continues by considering the nodes in the subsequent modules until the covering of the entire network.

e) Average node total control accuracy $\langle\delta\rangle + \langle\eta\rangle$ for single-drivers as a function of the target size $m$. Magenta/green curves correspond to results obtained by averaging the total accuracy for the drivers inside/outside the target. Results correspond to the mean values obtained from 100 simulated networks.

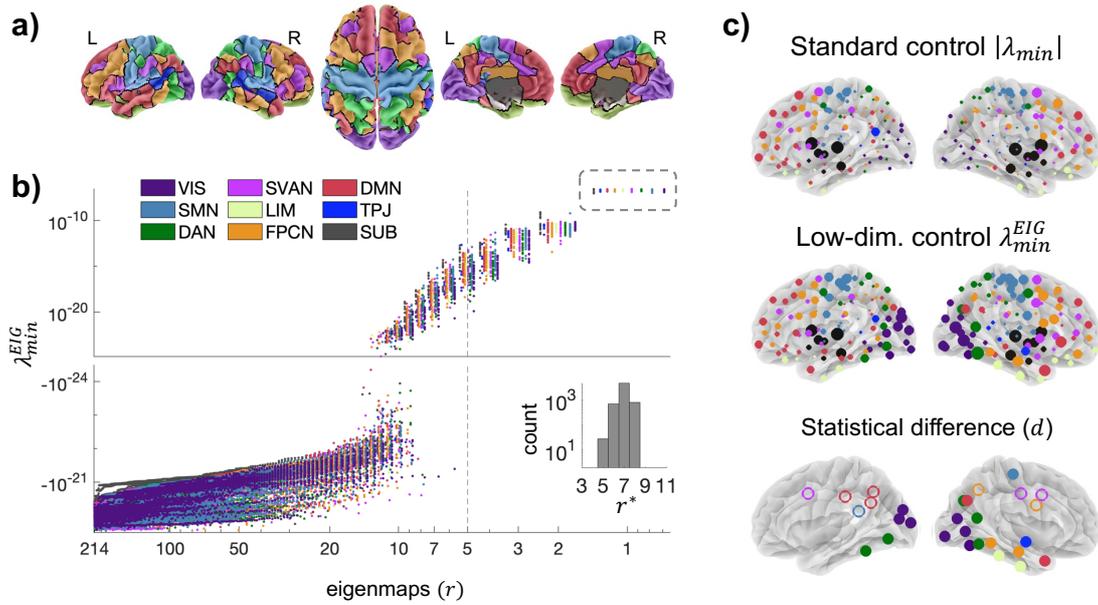

**Figure 3 - Single-driver controllability of brain networks.**

a) The Yeo2011 brain atlas parcellation. Each of the 214 regions of interests (ROIs) are organized in 9 functional systems: the visual network VIS, the somatomotor network SMN, the dorsal attention network DAN, the saliency and ventral attention network SVAN, the limbic network LIM, the frontoparietal control network FPCN, the default mode network DMN, the temporoparietal junction TPJ, and the subcortical network SUB.

b) Low-dimensional worst-case control centrality $\lambda_{min}^{EIG}$ values as a function of the number of eigenmaps $r$. Each point corresponds to a different node (ROI) controlling the entire brain. Colors code for different systems. Values are shown for a representative subject. By decreasing $r$ all $\lambda_{min}^{EIG}$ values become positive and numerically reliable after a critical threshold $r^*$. The inset illustrates the distribution of $r^*$ from all subjects (N=6134). Note that the standard metric $\lambda_{min}$ ($r = n = 214$) gives the lowest negative values making it difficult to interpret.

c) Group-averaged spatial distribution of standard ($|\lambda_{min}|$) and low-dimensional ($\lambda_{min}^{EIG}$) control centrality. Low-dimensional control centrality exhibits a significant reorganization compared to standard control centrality. The third row shows the ROIs that significantly gain (filled circles) or loose (empty circles) importance as compared to standard control (Sign test $p \ll 10^{-6}$, Cohen' $|d| > 0.5$, File S2).

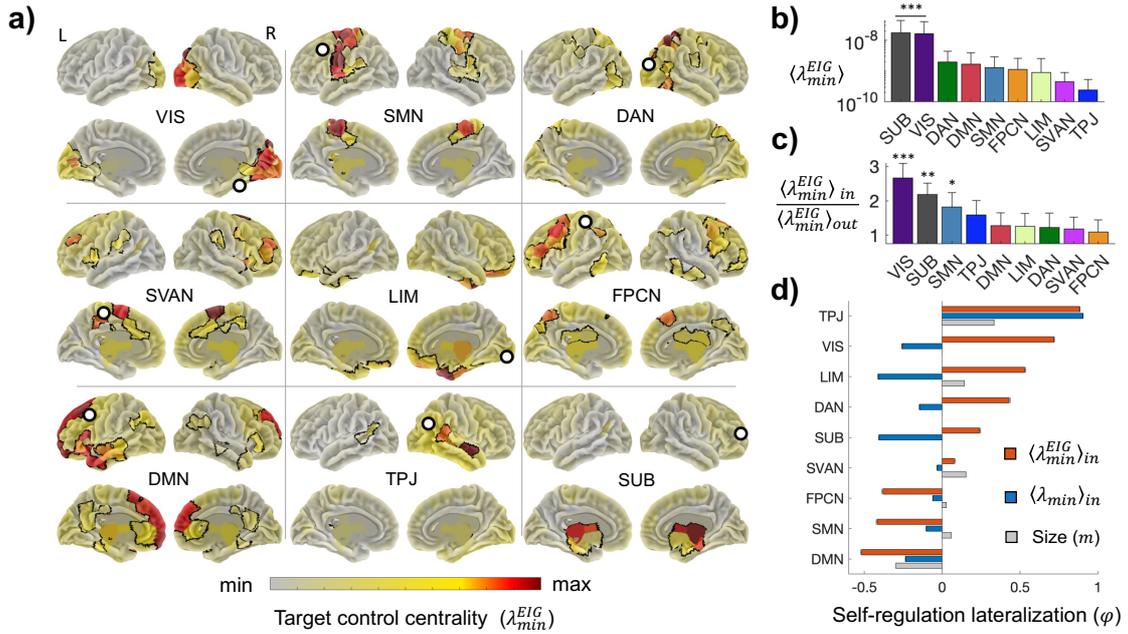

**Figure 4 - Target controllability of brain systems.**

a) Spatial distribution of group-averaged control centrality $\lambda_{min}^{EIG}$ when targeting each separate brain system. Target systems are contoured by black curves. Best drivers tend to fall within each target as indicated by the colorbar. White circles identify the best drivers outside the target (Tab 1). For each system results are illustrated for left (L) and right (R) hemisphere in both ventral (up) and dorsal (bottom) views.

b) System controllability $\langle \lambda_{min}^{EIG} \rangle$ as the mean control centrality of all the nodes targeting a specific system. Bars indicate group-averaged values and error bars standard deviations. Asterisks denote the systems whose controllability is significantly higher according to a post-hoc ANOVA analysis ($p \ll 10^{-6}$). The more the asterisks, the stronger the difference. Tukey-HSD Post-hoc ***$p \ll 10^{-6}$, **$p < 0.001$,*$p < 0.05$ (File S2).

c) Ratio between self-regulation $\langle \lambda_{min}^{EIG} \rangle_{in}$ and external regulation $\langle \lambda_{min}^{EIG} \rangle_{out}$ measured respectively by the mean control centrality of the nodes inside and outside a specific targeted system. Bars indicate group-averaged values and error bars standard deviations. Values have been log-transformed for the sake of readability. Asterisks denote the systems whose controllability ratio is significantly higher according to a post-hoc ANOVA analysis ($p \ll 10^{-6}$). The more the asterisks, the stronger the difference. Tukey-HSD Post-hoc ***$p \ll 10^{-6}$, **$p < 0.001$,*$p < 0.05$ (File S2).

d) System lateralization in terms of self-regulation from the right (R) and left (L) hemisphere $\varphi = \frac{\zeta^R - \zeta^L}{\zeta^R + \zeta^L}$. Red colors correspond to low-dimensional controllability $\zeta = \langle \lambda_{min}^{EIG} \rangle_{in}$. Blue colors correspond to standard controllability $\zeta = \langle \lambda_{min} \rangle_{in}$. Grey colors show the lateralization in terms of number of nodes of the systems in each hemisphere. Bars indicate group-averaged values and errorbars standard error means. Lateralization of low-dim. self-regulation significantly depends on the brain system (ANOVA, $p \ll 10^{-6}$).

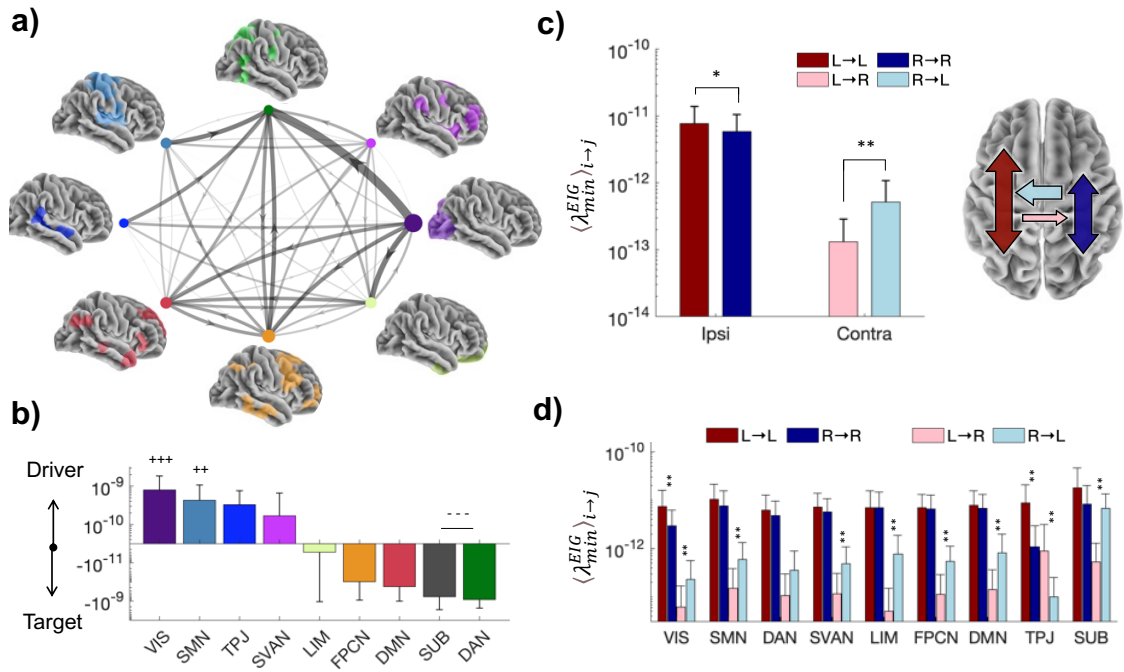

**Figure 5 – Control relationships between brain systems**

a) The group-averaged controllability meta-graph. Nodes correspond to different brain systems. Directed weighted links illustrate the group-averaged mean control centrality of system *i* when targeting system *j* $\langle \lambda_{min}^{EIG} \rangle_{i \to j}$. The darker and thicker the link, the stronger the influence of system *i* on *j*. Self-loops and the SUB network are not represented as their control centrality is several orders of magnitudes higher.

b) System control unbalance as the difference between the sum of outgoing and incoming weighted links from the individual meta-graphs. Positive values=tendency to act as driver. Negative value= tendency to act as target. Bars indicate group-averaged values and errorbars standard deviations. +/- denote the systems whose controllability is significantly higher/lower according to a post-hoc ANOVA analysis ($p \ll 10^{-6}$). The more the symbols, the stronger the differences. Tukey-HSD Post-hoc ***$p \ll 10^{-6}$, **$p < 0.001$ (File S2).

c) Hemispheric preference in terms of ipsilateral and contralateral control capacity. Dark colors denote *ipsilateral* control as the mean of the nodes in hemisphere *i* targeting the same hemisphere $\langle \lambda_{min}^{EIG} \rangle_{i \to i}$. Light colors indicate *contralateral* control as the mean of the nodes in hemisphere *i* targeting the other hemisphere $\langle \lambda_{min}^{EIG} \rangle_{i \to j}$. Bars indicate group-averaged values and errorbars standard deviations. L=left hemisphere, R=right hemisphere. Asterisks indicate to Cohen's *d* values measuring effect sizes. Sign test **$p \ll 10^{-6}$, Cohen' $|d| > 0.5$, *$p \ll 10^{-6}$, Cohen' $|d| > 0.2$, File S2)

d) Hemispheric preference of single systems in terms of their ability to control the entire ipsilateral and contralateral hemisphere. Same graphical conventions as before.

| Target system | Inside driver | Outside driver |
|---|---|---|
| VIS | VisPeri_ExStrSup_3-R | LimbicA_TempPole_4-R |
| SMN | SomMotB_Cent_1-L | ContB_PFCl_1-L |
| DAN | DorsAttnA_SPL_4-R | VisCent_ExStr_5-R |
| SVAN | SalVentAttnA_FrMed_2-R | SomMotA_7-L |
| LIM | LimbicA_TempPole_1-R | VisCent_Striate_1-R |
| FPCN | ContA_IPS_1-L | SomMotA_7-L |
| DMN | DefaultA_PFCd_1-L | ContB_PFCl_1-L |
| TPJ | TempPar_1-R | DefaultC_IPL_1-R |
| SUB | Sub_thalamusproper-R | DefaultA_PFCm_3-R |

**Table 1: Driver-target best configurations.**

Drivers are selected according to their maximum group-averaged low-dimensional control centrality $\lambda_{min}^{EIG}$.

## Supplementary text S1

**Optimal control: input signal derivation**
In the results section, we validated our framework for synthetic networks by simulating trajectories. We first, need to find the control input $u(t)$. To do so we solve the following optimization problem:

$$min_u \{ J_\rho(u, t_f) = \left(y_f - y(t_f)\right)^T \left(y_f - y(t_f)\right) + \rho \int_0^{t_f} u(\tau)^T u(\tau) \, d\tau \} \quad (1)$$
$$s.t. \quad \dot{x}(t) = Ax(t) + Bu(t); \; y(t) = Cx(t); \; x(0) = x_0$$

Overall, it is a minimization problem with soft constraints on the output. The term $\rho \int_0^{t_f} u(\tau)^T u(\tau) d\tau$, called the scalar running cost function, minimizes the energy and the term $E\left(x(t_f)\right) = \left(y_f - y(t_f)\right)^T \left(y_f - y(t_f)\right) = \left(y_f - Cx(t_f)\right)^T \left(y_f - Cx(t_f)\right)$, called the scalar terminal cost function, is where we express the final constraint on the output.

The problem can be solved using Pontriyargin's maximum principle[1] (see sections 5.1 and 5.2) by introducing the Hamiltonian equation:

$$\mathcal{H}(x(t), v(t), u(t)) = \rho \, u(t)^T u(t) + v(t)^T (Ax(t) + Bu(t)), \quad (2)$$

where $v(t)$ is the vector of the adjoint states.
We know from the fundamentals of optimal control theory[2] that the optimal trajectory $(x^*, v^*, u^*)$ is the solution to the following equations:

- Adjoint equation: $\frac{\partial \mathcal{H}}{\partial x} = \dot{v}(t) = -A^T v(t)$ (3)

- Stationary equation: $0 = \frac{\partial \mathcal{H}}{\partial u} = 2\rho u(t) + B^T v(t)$ (4)

- Boundary/Transversality condition: $v(t_f) = \frac{\partial E(x(t_f))}{\partial x(t_f)} = -2C^T(y_f - Cx(t_f))$ (5)

The adjoint and stationary equations, (3) and (4), can be rewritten in condensed form:

$$\begin{bmatrix} \dot{x}^* \\ \dot{v}^* \end{bmatrix} = \begin{bmatrix} A & -(2\rho)^{-1}BB^T \\ 0 & -A^T \end{bmatrix} \begin{bmatrix} x^* \\ v^* \end{bmatrix} = H \begin{bmatrix} x^* \\ v^* \end{bmatrix} \quad (6)$$

where $H$ is the Hamiltonian matrix. Equation (6) can be solved as:

$$\begin{bmatrix} x^*(t) \\ v^*(t) \end{bmatrix} = e^{tH} \begin{bmatrix} x^*(0) \\ v^*(0) \end{bmatrix} \quad (7)$$

This way, we obtained an expression for $v^*(t)$ and the problem is almost solved since equation (4) gives $u^*(t) = \frac{-B^T v(t)}{2\rho}$. We still are left with the unknown $v^*(0)$, which can be found using the boundary conditions[3]. If we note $e^{t_f H} = \begin{bmatrix} M_{11} & M_{12} \\ M_{21} & M_{22} \end{bmatrix}$, we have the three following boundary constraints:

$$\begin{cases} x^*(t_f) = M_{11} x_0 + M_{12} v^*(0) \\ v^*(t_f) = M_{21} x_0 + M_{22} v^*(0) \\ v^*(t_f) = -2C^T(y_f - Cx^*(t_f)) \end{cases} \quad (8)$$

The first two equations come from (7) and the third is the Hamiltonian boundary condition (5). We have three unknowns $v^*(0)$, $x^*(t_f)$, $v^*(t_f)$, and three equations. The calculus gives:

$$v^*(0) = (M_{22} - 2C^T C M_{12})^\dagger ((2C^T C M_{11} - M_{21})x_0 - 2C^T y_f), \qquad (9)$$

where † denotes the pseudo-inverse of a matrix.

**Laplacian of the network and state matrix**

For a graph with adjacency matrix $G$ the Laplacian matrix is defined as $L(G) = D_G - G$, where $D_G$ is the diagonal matrix containing the degree sequence of nodes of $G$. In this work, we stabilize the nodes' dynamics and consider the matrix $A = G - cI$, where $c$ is an arbitrary constant. We verify here, using the linearity of the Laplacian operator, that $L(A) = L(G)$.

$$L(A) = L(G - cI) = L(G) - L(cI) = (D_G - G) - (cI - cI) = D_G - G = L(G) \qquad (10)$$

**References**

1. Kirk, D. E. *Optimal Control Theory: An Introduction*. (Courier Corporation, 2012).

2. Anderson, B. D. O. & Moore, J. B. *Optimal Control: Linear Quadratic Methods*. (Prentice Hall, 1990).

3. Gu, S. *et al.* Optimal trajectories of brain state transitions. *NeuroImage* **148**, 305–317 (2017).

# Supplementary figures (S1-S5)

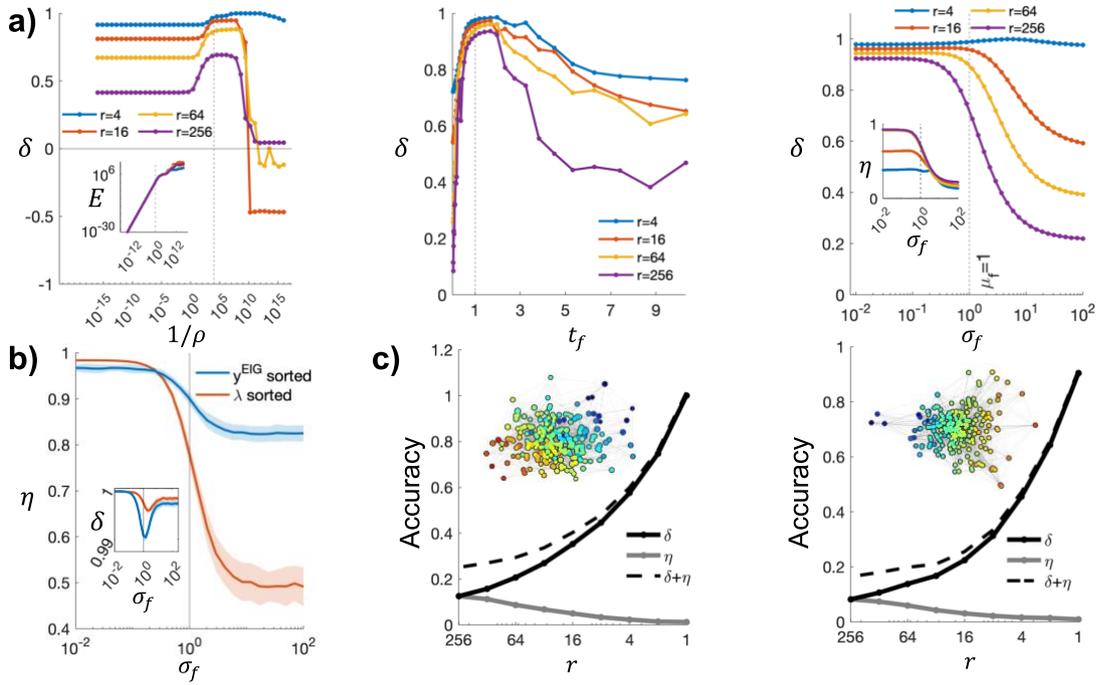

**Fig S1 – Effects of the parameters and topology**

a) Precision $\delta$ as a function of the simulation parameters: the reciprocal of the regularization parameter $1/\rho$, the final time $t_f$, and the final state dispersion $\sigma_f$, when controlling $r$ = 4, 16, 64, or 256 eigenmaps. The parameters are chosen within the ranges corresponding to good performance in terms of precision, i.e., $\rho$ =10$^{-4}$ and $t_f$ = 1. The final states $x_f$ were sampled from a Gaussian law with a fixed mean of $\mu_f$=1 and standard deviation $\sigma_f$. Highly dispersed final states were harder to attain ($r$=256) and controlling fewer eigenmaps resulted in better precision $\delta$ but worse representativeness $\eta$. We fixed $\sigma_f$=10 in the main results section to emphasize the role of the output dimension/number of eigenmaps.

b) Representativeness $\eta$ and precision $\delta$ (inset) as a function of the final state dispersion $\sigma_f$ using $r$=64 most representative eigenmaps sorted according to the magnitude of the eigenstate (in blue) or using the $r$ first eigenmaps corresponding to the smaller Laplacian eigenvalues (in red). We used all nodes as drivers $n_d$=$n$ and the final state had a fixed mean of $\mu_f$=1. for smooth final states with small small dispersion, both eigenmaps selection schemes gave high and similar representativeness $\eta$ with a slight superiority of the $\lambda$-sorted scheme. As the dispersion increased, $\sigma_f$>1, selecting the eigenmaps accordingly to the eigenstate gave better representativeness $\eta$ and that was accompanied with lower precision $\delta$ in the low-diensional space as the control task was harder. Results are averaged over 100 HMSW network realizations and shaded areas represent standard deviations.

c) Precision $\delta$, representativeness $\eta$, and their sum as a function of the number of controlled eigenmaps $r$ for two different topologies: the Erdos-Renyi model ER($n$=256 , $p$=0.035), and the Barabasi-Albert model BA($n$=256 and connection density 0.035, bias $\gamma$= 2). d de. We also the same control task as for the HMSW model: the final state was sampled from $\mathcal{N}(\mu_f$=1, $\sigma_f^2$=100), and the trajectory was simulated for $\rho$=10$^{-4}$ and $t_f$=1. Results are averaged over 100 network realizations.

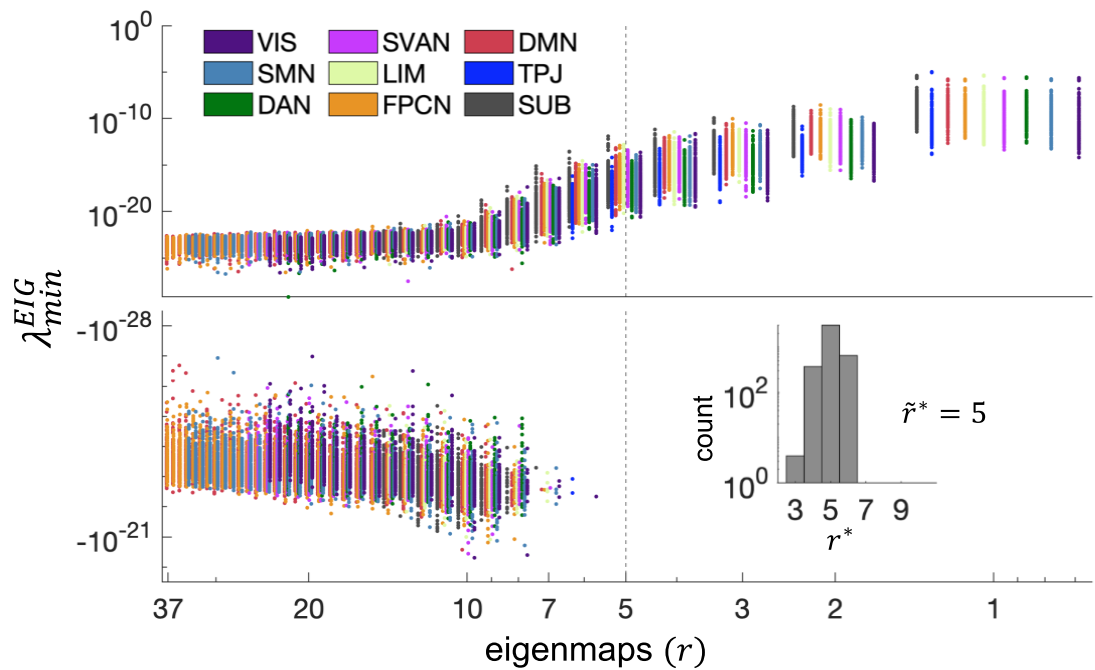

**Fig S2 - Single-driver controllability of target brain networks: effect of dimension.**

Low-dimensional worst-case control centrality $\lambda_{min}^{EIG}$ values as a function of the number of eigenmaps $r$. For each of the 9 columns and colors, a different brain system is taken as target and each point corresponds to a different node (ROI). Values are shown for a representative subject. By decreasing $r$, all $\lambda_{min}^{EIG}$ values become positive and numerically reliable after a critical threshold $r^*$. The inset illustrates the distribution of $r^*$ from all subjects (N=6134). The group median $\tilde{r}^* = 5$ has been chosen as representative value for each subject. Note that the $\lambda_{min}^{EIG}$ is equivalent to the standard metric $\lambda_{min}$ when $r = m$.

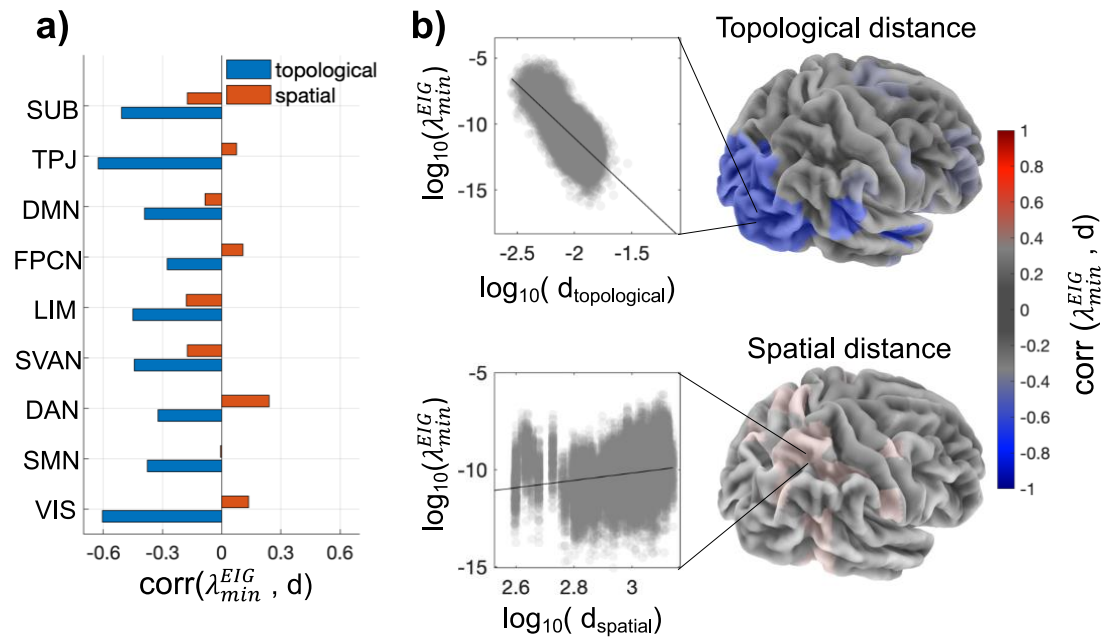

**Fig S3 - Relationship between low-dimensional controllability and distance metrics.**

a) Pearson correlation between low-dimensional controllability and distances between drivers and targeted networks. We considered the distance to the targeted network as the sum of the distances to its nodes $d_{i,\ S_{net}} = \sum_{j \in S_{net}} d_{i,j}$. Topological (blue) refers to the length of the shortest path and spatial (red) to the Euclidean distance.

b) Visualization of the correlation coefficients and scatter plots for two representative brain systems, i.e. the VIS network regarding topological distance and the DAN regarding spatial distance.

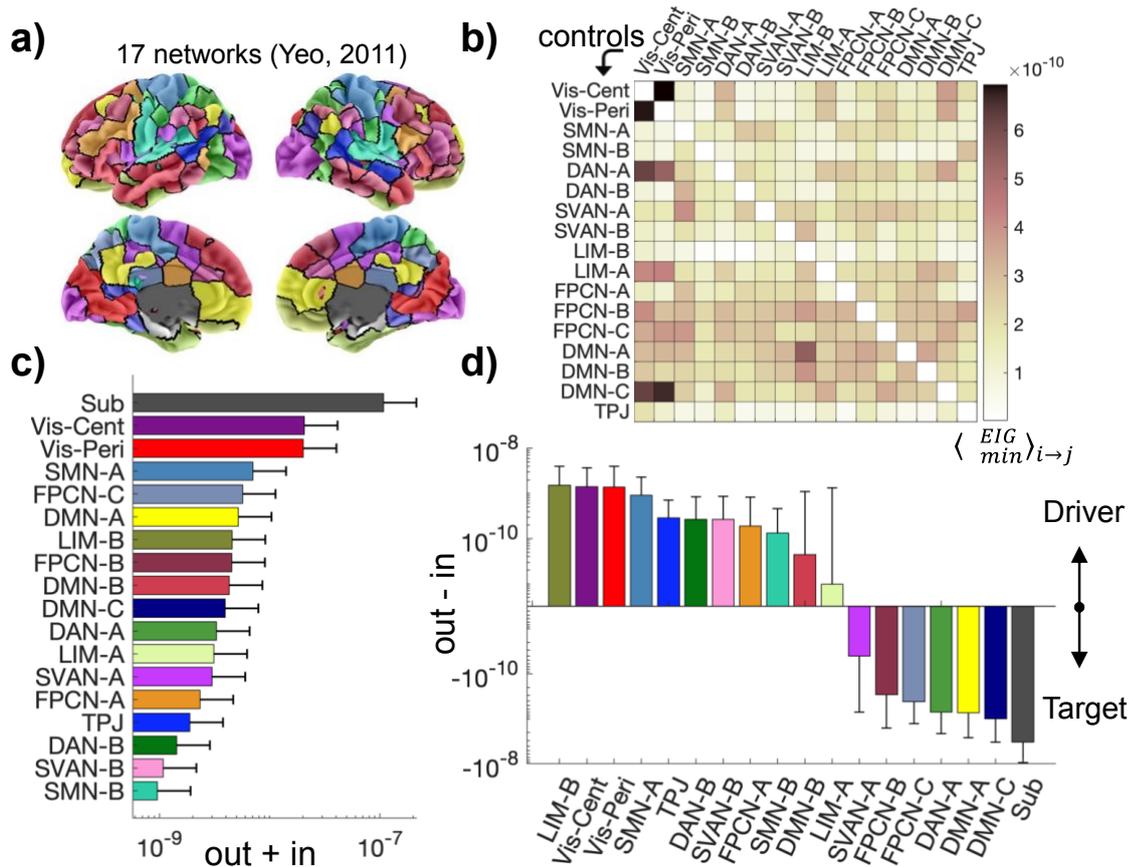

**Fig S4 - Control relationships between brain systems in a finer parcellation.**

a) The 17 split components of the Yeo2011 brain atlas parcellation.

b) Adjacency matrix of the group-averaged controllability meta-graph between the different brain systems. The links represented by color are the group-average of control centrality of the system in column *i* when targeting system in line *j* $\langle \lambda_{min}^{EIG} \rangle_{i \to j}$. $\langle \lambda_{min}^{EIG} \rangle_{i \to j}$ is obtained by taking the geometric mean control centrality $\lambda_{min}^{EIG}$ of drivers in system *i* when targeting system in *j*. Self-loops and the SUB network are not represented as their control centrality is several orders of magnitudes higher.

c) System total contribution as the sum of outgoing and incoming weighted links from the individual meta-graphs. Bars indicate group-average values and error bars standard deviations.

d) System control unbalance as the difference between the sum of outgoing and incoming weighted links from the individual meta-graphs. Positive values=tendency to act as driver. Negative value= tendency to act as target. Bars indicate group-average values and error bars standard deviations.

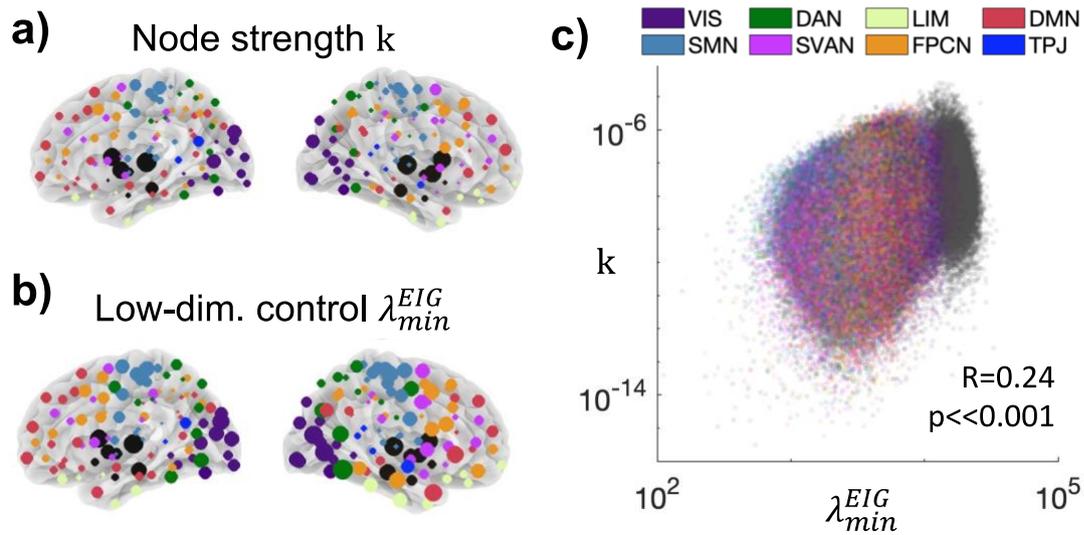

**Fig S5 - Relationship between low-dimensional controllability and node strength.**

a) Group-averaged spatial distribution of node strength $k$ given by the sum of all the weighted links of a node.

b) Group-averaged spatial distribution of low-dimensional ($\lambda_{min}^{EIG}$) control centrality.

c) Scatter plot of node strength $k$ and low-dimensional ($\lambda_{min}^{EIG}$) control centrality for all $n \times N$ nodes and participants. Pearson correlation test revealed a low correlation between the two metrics.